# Computed Tomography imaging of large cargos

Polad M. Shikhaliev

*Applied Physics Initiatives, LLC, Baton Rouge, LA 70879*

A B S T R A C T

A feasibility study of the megavoltage (MV) computed tomography (CT) for imaging large cargos has been performed. The MV cargo CT system has imaging field of view of 3.25m in diameter. The system design includes a compact x-ray source and detectors that are already used in existing MV cargo radiography systems. Although compact x-ray sources provide limited x-ray output, it is shown that this output is sufficient to perform CT imaging of large cluttered cargo contents for which MV radiography may be inconclusive. The scan time is 60sec. at the x-ray dose levels lower than the permitted limit of 5mGy. A steel tank phantom with 2m diameter including different materials was used. The MV CT images were generated at 3.5MV, 6MV, and 9MV beam energies. The materials were separated based on their CT numbers as well as using dual energy subtraction method. The materials with close densities and x-ray attenuations such as W and U were reliably separated and quantified based on their measured CT numbers. The boundary conditions for the system design were determined including scan time, radiation dose, maximum object thicknesses, and spatial resolution. Based on the findings of this feasibility study, it is concluded that the MV cargo CT system with large field of view can be developed based on existing technology already used in MV cargo radiography. Such a CT system can be extremely useful as adjunct to existing MV radiography on resolving the special cases that cannot be resolved conclusively with radiography alone.

**Key words:** cargo imaging, computed tomography, dual energy, material decomposition

*E-mail address:* psm4165@gmail.com

## 1. Introduction

Hundreds of millions of large cargo transactions are performed annually worldwide. These cargos can potentially be used for smuggling illicit materials including explosives, narcotics, nuclear materials, shielded radiation sources, etc. The large cargos may have sizes of up to few meters in cross sections and average densities of about thousand kilograms per cubic meter. They may have cluttered arrangements of the contents which can serve appropriate environment for concealing illicit materials. Routine inspections and security screenings are necessary to detect these illicit activities and x-ray imaging is one of the widely used non-intrusive screening methods. The high-energy x-ray beams with high penetration capabilities are required for imaging large cargos. The x-rays with mega-electronvolt (MeV) energies are appropriate for this purpose and corresponding mega-voltage (MV) cargo radiography systems have been developed, investigated [1-11] and are commercially available from several companies [12-16].

Although MV radiography is the method of choice for fast and non-intrusive imaging of large cargos, this method has a major limitation related to image overlapping. Because radiography image is a projection of the 3D structure of the object to a 2D image plane, the images of the components are overlapped. Therefore, relatively small and/or low contrast objects can be lost over the cluttered image background of the overlapped structures. One possibility for improving the overlap problem is taking a second image at different view angle with respect to the first image. This can provide additional information about the cargo content that was not available in the first image. However, the second image itself may suffer from image overlap and it may not resolve the problem. Another method is using manual investigation when the content of the cargo is removed and manually investigated. However, this method is time and labor consuming, and expensive. Also, additional formal requirements may be imposed due to the intrusive nature of manual screening.

The image overlap problem of projection radiography is well known in many applications, including medical imaging, airport baggage inspection, etc. Most effective solution to this problem is computed tomography (CT) imaging which provides true 3D imaging of internal structures of the objects. However, megavoltage CT of large cargo has not been feasible so far due to a number of constraints including large sizes of the required CT gantry and heavy weight of the MV x-rays source which has to be rotated around the imaged object.

In the current work we propose and investigate a large scale MV CT system for imaging large cargos. Comprehensive feasibility studies were performed using the boundary conditions for the system parameters determined by current state of the art of existing technologies. The MV cargo CT system has 4m gantry diameter and 3.25m useful field of view. The system uses a portable MV x-ray source that operates in 2-9MV range and provides sufficient x-ray power to perform a single- or few-slice CT acquisition in 60 seconds scan time and at the x-ray dose levels comparable to that used in MV radiography. Several large size cargo phantoms were used for quantitative and qualitative assessments of the system performance. It appears that the described MV cargo CT can be adjunct to MV radiography. It can be extremely helpful in cases when MV radiography is not conclusive due to the image overlap and clutter problems, and the MV CT acquisition can provide detailed cross-sectional image of the concerned areas.

## 2. Methods and materials

### 2.1. System design

The design concept of the MV cargo CT system is shown in **figure 1**. The key components of the system include MV x-ray source and curved array of the x-ray detectors. The x-ray source and detector array are conceptually similar to that used in MV radiography except that they can rotate around the object to acquire CT data. The detector array is arranged along the circular arc to minimize the CT gantry size. The system has 4m source-to-detector distance, 2m source-to-isocenter distance, and magnification factor of 2. The above geometrical parameters were selected to provide useful field of view of 3.25m which is sufficient for imaging large cargos, including containers with $2.3 \times 2.3 m^2$ standard cross-sectional sizes.

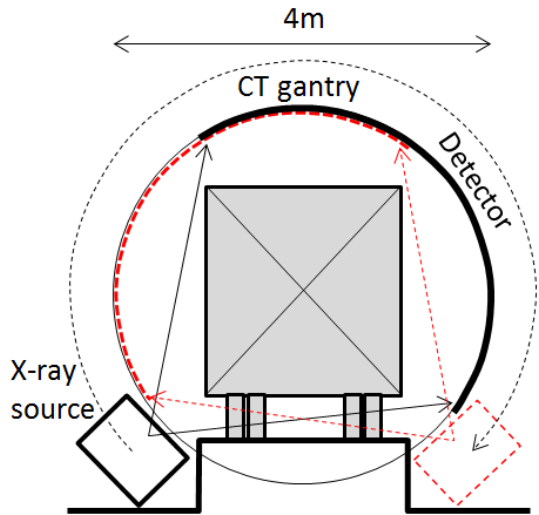

Figure 1. Schematics of the large scale MV cargo CT system

Another key feature of the MV cargo CT design is that the CT data is acquired in less than 360 degree gantry rotation angle. This is possible because CT reconstruction requires minimum scanning angle that equals to (180+fan angle) degree [17]. The fan angle in our design is 78 degree, thus 258 degree gantry rotation is sufficient for a single-slice CT acquisition. This is important because 360 degree gantry rotation would be impractical due to the large size of the x-ray source. On the other hand, the detector array has a compact cross-sectional dimensions and it can move underneath of the cargo through a narrow slot. Notice that the system can perform also single- and multiple-view radiography images in addition to CT acquisition.

One of the key requirements for feasibility of the MV cargo CT is that the x-ray source should have a limited weight and compact size, and at the same time it should provide sufficiently high x-ray output. The MV x-ray source should operate at up to 9MV beam energies and be compact enough to be mounted on the CT gantry with 4m diameter which should rotate around the cargo during CT acquisition. It is known that the MV x-ray sources with high x-ray outputs are heavy and large in size. For example, the 6/9MV linear accelerator (LINAC) based source with 30Gy/min dose rate at 1m used in cargo radiography [8, 18] have approximately $1.6 \times 1.0 \times 0.6 m^3$ sizes and 1500kg weight [18]. Such a source would be suboptimal for MV cargo CT applications. The other commercially available mobile MV radiography systems use more compact sources but their specifications are not disclosed [13]. In the current work we used published specifications of the Betatron based MV x-ray source which provides up to 9MV x-ray energies, 0.2Gy/min dose rate at 1m, total weight of 260kg, and cylindrical radiator with 45cm diameter and 34cm height. Although the x-ray dose rate 0.2Gy/min of this source is much lower than that of above LINAC based system, it is sufficient to perform CT scans as will be shown below in section 2.2.

**2.2. CT acquisition parameters**

The CT acquisition parameters are determined by several boundary conditions. First, the radiation dose applied to the cargo should be limited to 500mRem (which is equivalent to 5mGy air kerma) because of possibilities of the stowaways in the cargo [6]. Second, the spatial resolution of the MV cargo CT should be comparable to that established in MV cargo radiography. Third, the output power of the x-ray source should be sufficiently high so that the CT scan is completed within practically acceptable time. The above boundary conditions determine the image quality in terms of signal-to-noise ratio and spatial resolution, which in turn, determine whether or not the MV Cargo CT could be useful for practical applications.

In the current study, we used the MV x-ray source with the electron energies of 3.5MeV, 6MeV, and 9MeV which determine the endpoint energies of the corresponding x-ray spectra [19]. The x-ray spectra were filtered with 0.5mm Pb filter to remove the low-energy photons from the spectra (**figure 2**). The low-energy photons should be removed because they do not contribute substantially to the image quality as they are almost fully absorbed in the imaged object. On the other hand, they contribute to the absorbed radiation dose which is not desirable.

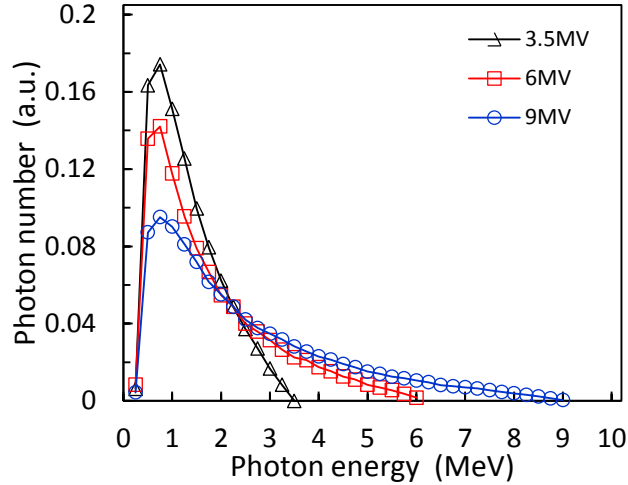

Figure 2. The x-ray energy spectra used for MV cargo CT acquisition.

For monoenergetic x-rays, the number $N$ of the x-ray photons per unit area per second at given energy $E$ and dose (air kerma) rate $D$ are related as

$$D = kN \left( \frac{\mu(E)}{\rho} \right)_{air} E \qquad (1)$$

where $\left( \frac{\mu(E)}{\rho} \right)_{air}$ is the mass-energy absorption coefficient of the air, and $k$ is the known constant [20]. For the polyenergetic x-ray beam passed through the filter material with thickness $t$ the dose rate is determined as

$$D = \int_{E_1}^{E_2} kN(E) e^{-\mu(E)t} \left( \frac{\mu(E)}{\rho} \right)_{air} E dE \qquad (2)$$

where $N(E)$ is the photon energy spectrum, and $E_1$ and $E_2$ are the lowest and highest photon energies in the spectrum, respectively. The x-ray dose output of the source 0.2Gy/min at 1m was converted to the photon numbers per mm$^2$ per second at 1m using the expressions (1) and (2). The total CT acquisition time (for 258 degree gantry rotation) was 60sec. which results in 15cm/sec linear speed of the source and detector along the circular gantry. This speed appears to be mechanically reasonable, and the 60sec. scan time also appears to be reasonable for secondary inspections. Also, 60sec. scan time satisfies the dose requirements as will be shown below.

The total number of CT projections was 720 which resulted in 0.0833sec. exposure time per CT projection. The photon numbers per pixel per CT projection were determined using the exposure time per CT projection, photon numbers per $mm^2$ per second from the source, the source to detector distance, and the detector pixel size of $4x4mm^2$ in the array and axial directions, respectively. The photon numbers per pixel per CT projection were randomized as Poisson distributed random numbers for reconstruction of realistic CT images that include image noise associated with photon statistics and radiation dose.

The radiation dose limit of 5mGy for MV cargo radiography was established with assumption that there may be illicit human transport (stowaways) within the cargo and that the whole body irradiation of the human may take place. However, in the MV cargo CT described in this work a single CT slice with 2mm slice thickness is irradiated, and the effective object dose is determined by averaging the single slice dose over the area of the object. The 60sec. CT acquisition time at 0.2Gy/min dose rate results in 0.2Gy dose absorbed within 2mm thick slice. Averaging this dose over the human body area with 50cm width gives 0.8mGy average dose, which is approximately by a factor of 6 lower than the dose limit of 5mGy. Therefore, there is a possibility of further increasing the dose that would improve the detected photon statistics and signal-to-noise ratio. This can be achieved by increasing the scan time, increasing CT slice thickness, or increasing both at a time, while maintaining the average absorbed dose below 5mGy limit.

The x-ray detector used in MV cargo CT system is conceptually similar to that used in MV cargo radiography [3, 9, 19]. The MV cargo CT design used in this study includes the array of 1360 pixels with pixel size of 4mm measured in both fan angle (pixel array) and axial (perpendicular to slice) directions. The detector pixels represent $CdWO_4$ scintillator crystals with 2cm thickness connected to a photodiode array. Taking into account the magnification factor of 2, the pixel size projected to the isocenter equals to 2mm, which determines the spatial resolution of the CT images. The detector operates in energy integrating mode and the signal response to the arrived photons is determined as:

$$S_i = \int_{E_1}^{E_2} N_D^i(E)\left(1 - e^{-\mu_D(E)t_D}\right) E \frac{\mu_D^{en}(E)}{\mu_D(E)} dE \qquad (3)$$

where $S_i$ is the signal generated in the pixel $i$, $\mu_D(E)$ and $\mu_D^{en}(E)$ are the total and energy attenuation coefficients, respectively, and $t_D$ is the detector thickness. In the expression (3) the approximation was made that the secondary photons (Compton scattered and annihilation photons) leave the detector pixel after primary photon interaction while the generated particles (Compton electrons and electron-positron pairs) deposit their energies in the pixel [3]. The estimations show that this is generally a fair approximation taking into account that the pixel size $4x4x20mm^3$ is small enough compared to the free path length of the high energy photons [21], and at the same time is large enough compared to the secondary particles ranges [22].

**2.3. Cargo phantoms**

While the external sizes and approximate average densities of the cargo containers are known, the types, shapes, and arrangements of the materials inside the cargo container can vary in wide ranges. Therefore, it is difficult to design a standardized MV cargo phantom that could be sufficiently representative for most cargos. In this feasibility study, we designed and used several cargo phantoms for quantitative and qualitative assessments. The phantoms included low-, medium-, and high-Z materials, including also material pairs with close linear attenuation coefficients (LAC) as shown in **figure 3**.

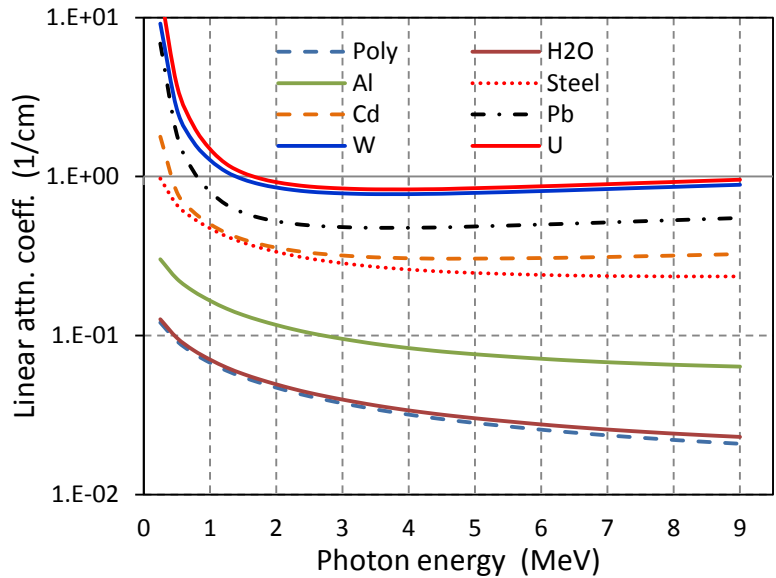

Figure 3. Linear attenuation coefficients of the materials used in cargo phantoms

The quantitative cargo phantom represents a cylindrical stainless steel tank having 2m diameter and 0.7cm wall thickness (**figure 4**). The cylindrical shape of the tank was chosen for consistency of quantitative analysis of the CT images while the wall material and thickness represents the commercial cargo tanks of similar sizes. The phantom included various materials of interests.

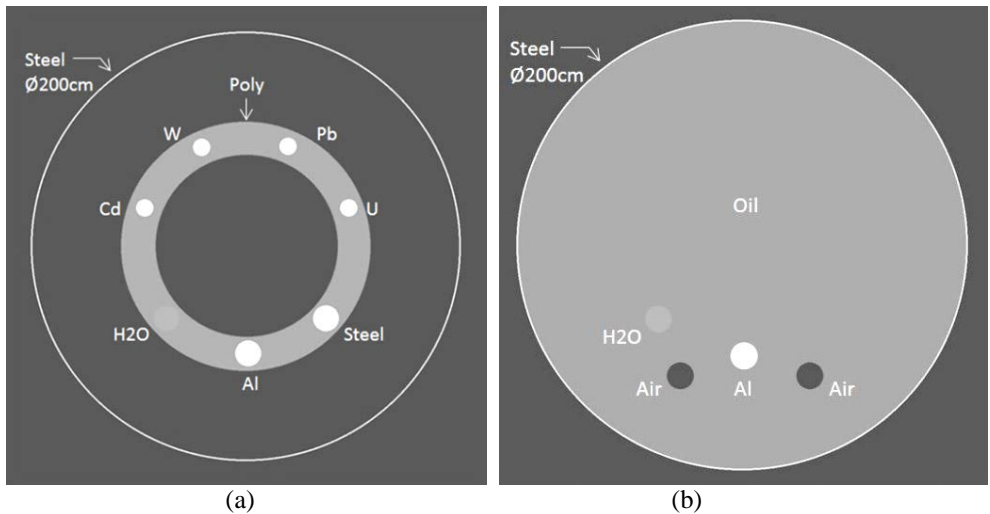

Figure 4. The cargo phantoms for quantitative MV CT imaging.

In one case the materials of interests were low-Z ($H_2O$, Al), medium-Z (steel, Cd), and high-Z (W, Pb, U) spheres embedded in the Polyethylene ring (**figure 4a**). The low- and medium-Z materials had 12cm diameters, and the high-Z materials had 8cm diameters. In another phantom design, the stainless steel tank was filled with oil, the medium-Z and high-Z materials were removed and only low-Z materials $H_2O$, Al, and air were included (**figure 4b**).

The qualitative cargo phantoms were designed to simulate clutter images to compare MV radiography and MV CT images in terms of image overlap (**figure 5**).

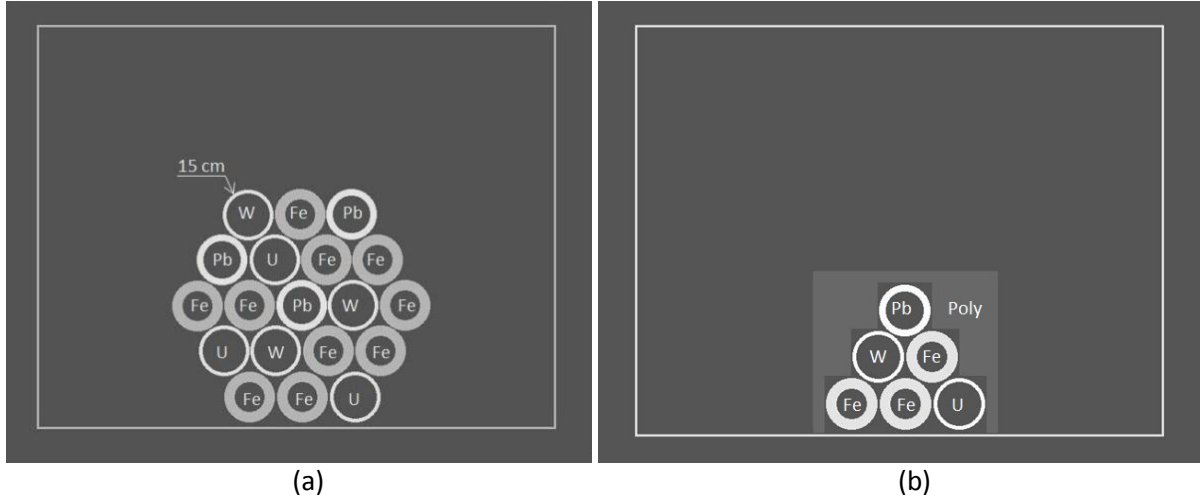

(a)                      (b)

Figure 5. The clutter phantom for qualitative imaging including a pile of 19 (a) and 6 (b) hollow cylindrical elements made from medium- and high-Z materials.

The phantoms represent a stainless steel container with $1.0 \times 1.25 m^2$ cross-sectional size and 0.7cm wall thickness, including a pile of 19 hollow cylinders with 15cm diameters (**figure 5a**). The cylinders were made of Fe, W, Pb, and U, and have wall thicknesses of 3.1cm, 1.1cm, 2cm, and 1.1cm, respectively. The wall thicknesses of the cylinders were selected such that their x-ray attenuations were comparable. In another setup the phantom includes 6 hollow cylinders covered by a Polyethylene cap (**figure 5b**). The second phantom configuration with 6 cylinders was used to demonstrate how the degree of x-ray attenuation affects the image quality.

**2.4. Data processing**

The CT images were reconstructed for 3.5MV, 6MV, and 9MV beam energies, at 0.8mGy total radiation dose (air kerma) measured at 1m for all images. The CT reconstruction was performed by filtered backprojection reconstruction method using 720 CT projections. The images reconstructed at 6MV and 9MV beam energies were used also for material decomposition with dual energy subtraction when one of the two materials is cancelled out from CT image. The details of the generation of CT projections and image reconstruction, as well material decomposition using dual-energy subtraction were reported in previous works [23-25].

The reconstructed CT images represent 2D distribution of the LAC across the cross-section of the object. Therefore, all materials can be physically separated and quantified based on cross-sectional CT images without overlap. The pixel values in reconstructed CT images are inherently relative values and do not equal to LAC. However, they are linearly transformed representations of the LAC. Therefore, in practice one uses the CT image of the calibration phantom with known elements to convert the relative pixel values to LAC in the object of interest.

Because the LAC are typically small values <1, it is not convenient to use them in practical applications. For example, in medical applications the LAC are linearly transformed to so-called "CT numbers" (also called "Hounsfield units") that are whole numbers [26]. This is performed by linear transformation of the relative LAC as:

$$CT\ number = 1000 \times \frac{\mu_{ij} - \mu_0}{\mu_0} \quad (4)$$

where $\mu_{ij}$ is the pixel value of the relative LAC for the pixel $(i, j)$, and $\mu_0$ is the LAC of the reference material. In medical applications, the reference material is selected to be water which is the main component of the human body. With water being a reference material, the CT numbers in CT images of the human body range from -1000 for air (LAC=0) to 3000 for bone (LAC=max), being zero for water. In analogy with medical applications, we introduced CT number concept in MV cargo CT with the reference material being steel, and corresponding CT numbers called "steel units" (SU). The SU in 9MV CT images range from -1000 for air to 2300 for uranium, being zero for steel. If we used another reference material, such as Al, then the CT numbers would range between -1000 and 9690, and for Cd the range would be between -1000 and 1710. Thus, the steel-based reference material seems to be optimal.

**3. Results**

The CT image of the cargo phantom including high-, medium-, and low-Z materials are shown in **figure 6a**. Corresponding image of the phantom filled with oil and including low-Z materials is shown in **figure 6b**. The image profiles across the horizontal line passing through the center of the phantom are shown in **figures 6c,d**, respectively.

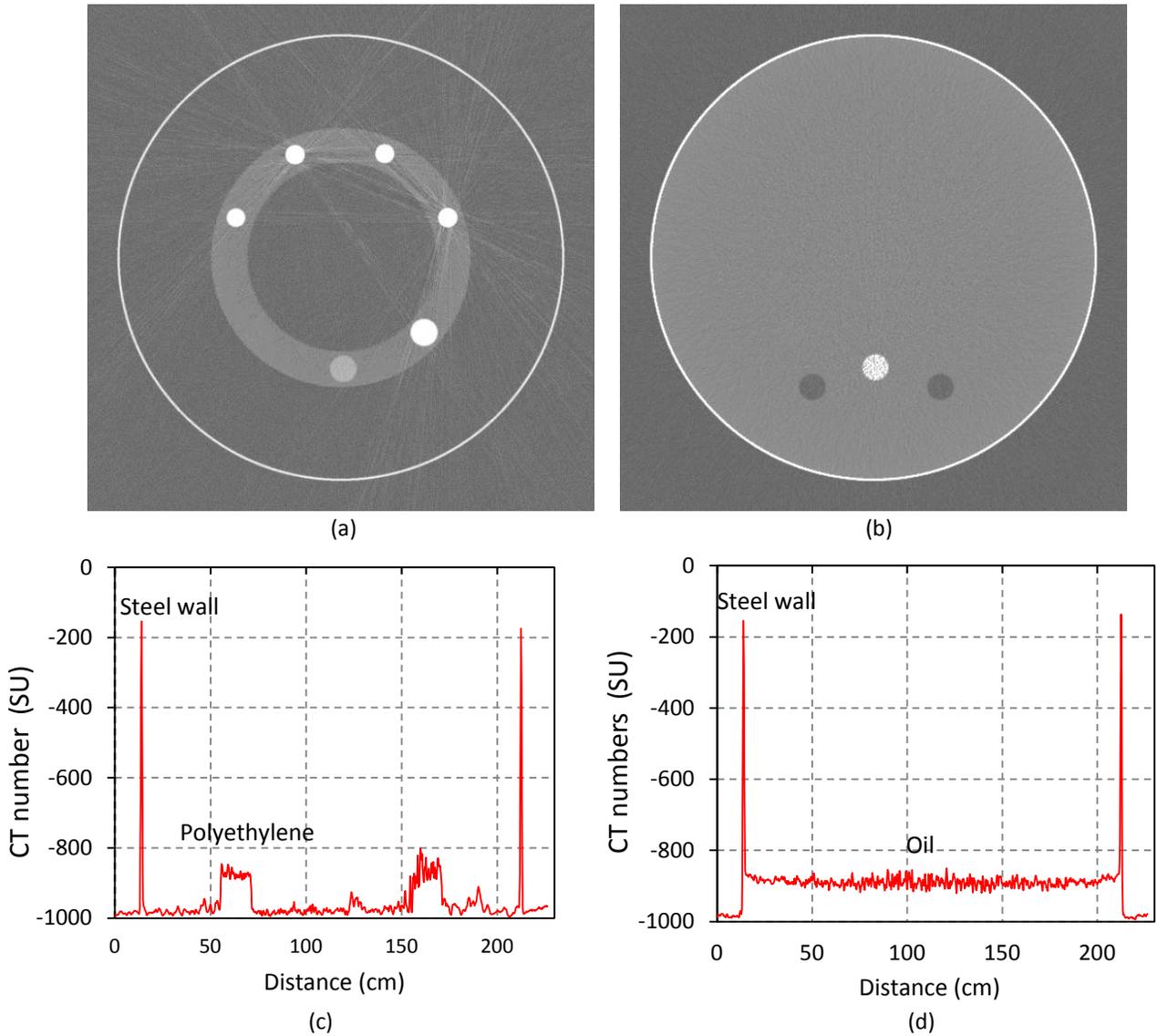

Figure 6. MV CT images of the steel tank phantom with 2m diameter including low-, medium-, and high-Z materials (a). Image of the same tank filled with oil and including H$_2$O (invisible) and Al spheres (b). Corresponding image profiles across the center of the phantoms (c, d), respectively.

The pixel values are given in CT numbers expressed in "steel units" (SU). As can be noticed, the image with high-Z materials exhibits major streak artifacts. These artifacts are associated with photon starvation effect due to strong attenuation of the x-ray photons by high-Z materials. The streaks are particularly noticeable across the lines passing through two high-Z materials at a time where the photon attenuation is highest. For the oil-filled phantom, although streak artifacts are absent, there is beam hardening (so called "cupping") artifacts where the CT numbers of oil is higher toward the edge of the phantom. Also, the image noise is higher toward the center of the phantom because the x-rays passing through the center exhibit stronger attenuation increasing statistical noise.

**Figure 7** shows the CT image profiles of the high-Z materials to demonstrate how well the Pb, W, and U can be quantified and separated one from another based on measured CT numbers. The profiles were plotted across the centers of the U and Pb (**figure 7a**) and U and W spheres (**figure 7b**). As can be seen, U can easily be separated from Pb based on CT numbers. Also, U can be separated from W despite the fact that the linear attenuation coefficients of the two materials are very close to each other.

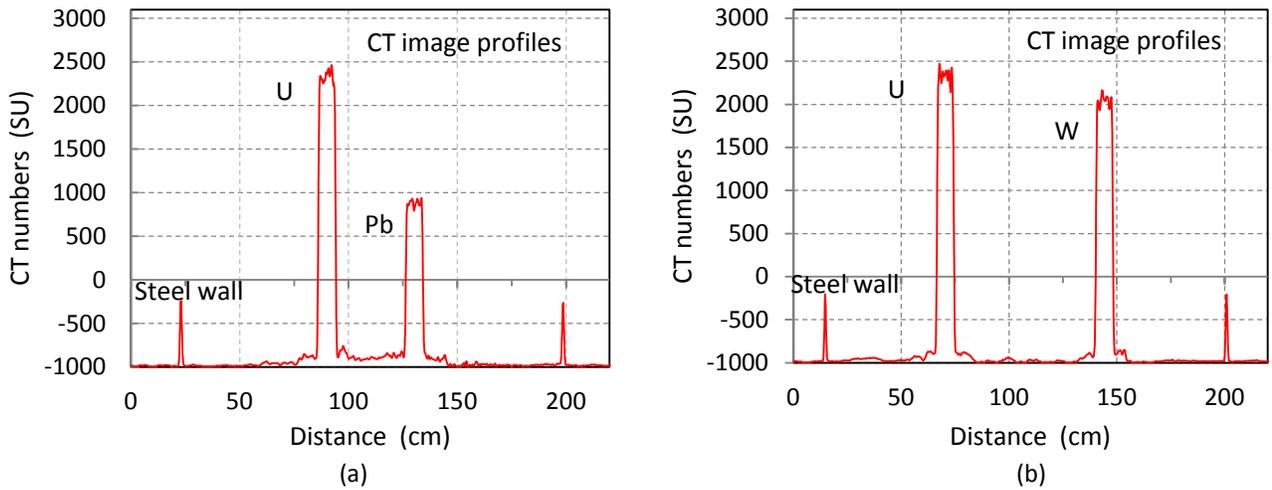

(a)                  (b)

Figure 7. Image profiles of the U and Pb (a) and U and W (b) spheres demonstrating how U can be separated from Pb and W based on their CT numbers expressed in steel units (SU).

**Figure 8** shows the trend of the CT numbers of low-, medium-, and high-Z materials in CT images acquired at 3.5MV, 6MV, and 9MV beam energies. As can be seen from **figure 8a**, water and Polyethylene are indistinguishable at all beam energies. On the other hand, U and W can be separated similarly at all three beam energies (**figure 8b**), although higher energy images exhibit lower noise due to higher x-ray penetration capabilities.

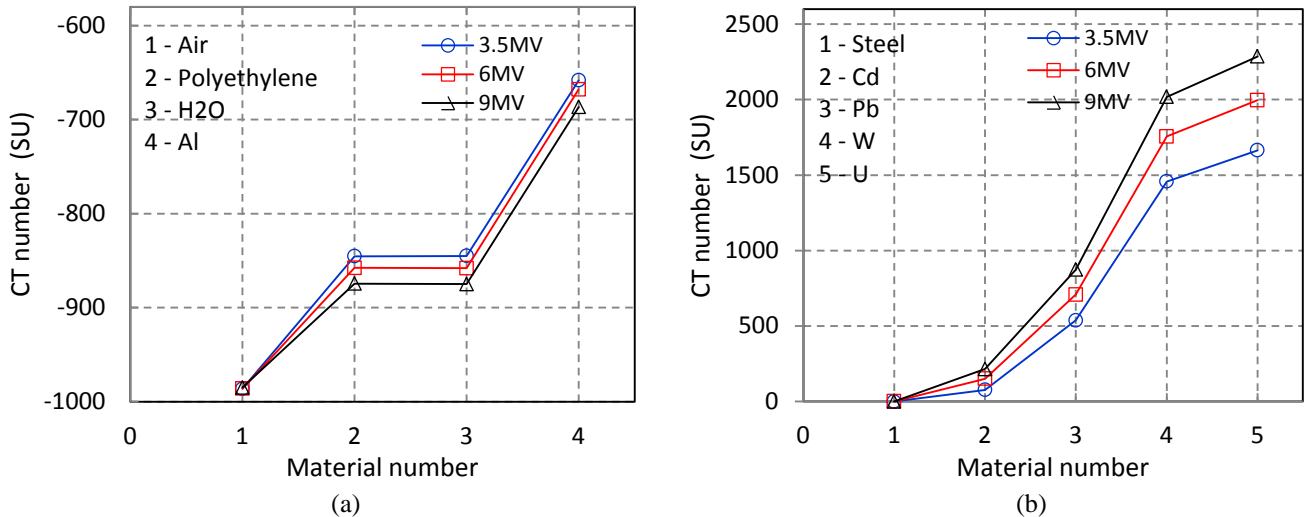

(a)                  (b)

Figure 8. Linear attenuation coefficients of low-Z (a) and medium/high-Z materials expressed in steel units (SU). The high-Z materials such as Pb, W, and U could be differentiated based on their SU.

The **Table 1** summarizes the mean values of the CT numbers and corresponding standard deviations for all materials and all beam energies. It is noticeable that at 9MV beam energy the CT number difference between U and W is by a factor of 6 larger than the standard deviations of CT numbers, which indicates that the two materials can be reliably separated based on measured CT numbers.

Table 1. CT numbers of and corresponding standard deviations

| Materials | 9MV | | 6MV | | 3MV | |
|---|---|---|---|---|---|---|
| | mean | std | mean | std | mean | std |
| U | 2298 | 49 | 1985 | 35 | 1641 | 105 |
| W | 2040 | 30 | 1757 | 56 | 1426 | 68 |
| Pb | 883 | 29 | 704 | 22 | 531 | 27 |
| Cd | 218 | 36 | 149 | 31 | 84 | 19 |
| Steel | -4 | 15 | -3 | 22 | -8 | 19 |
| Al | -688 | 6 | -663 | 17 | -659 | 12 |
| Polyethylene | -877 | 8 | -859 | 12 | -849 | 9 |

The high-Z materials with larger thicknesses attenuate and harden the x-ray beam stronger than the thinner materials. Therefore, the CT numbers of the same material may vary depending on the thickness of the material due to varying beam hardening. To test this effect, a CT phantom including U spheres with diameters 1cm, 2cm, 4cm, and 8cm was imaged (**figure 9**). This is the same phantom shown in **figure 2a** in which the elements Cd, W, and Pb were substituted with U spheres with 1cm, 2cm, and 4cm diameters, respectively. The magnitudes of the streak artifacts decrease with decreasing diameters of U spheres. Also, the image profiles of the U spheres show that the differences in CT numbers of the U spheres with different diameters are comparable to the standard deviations.

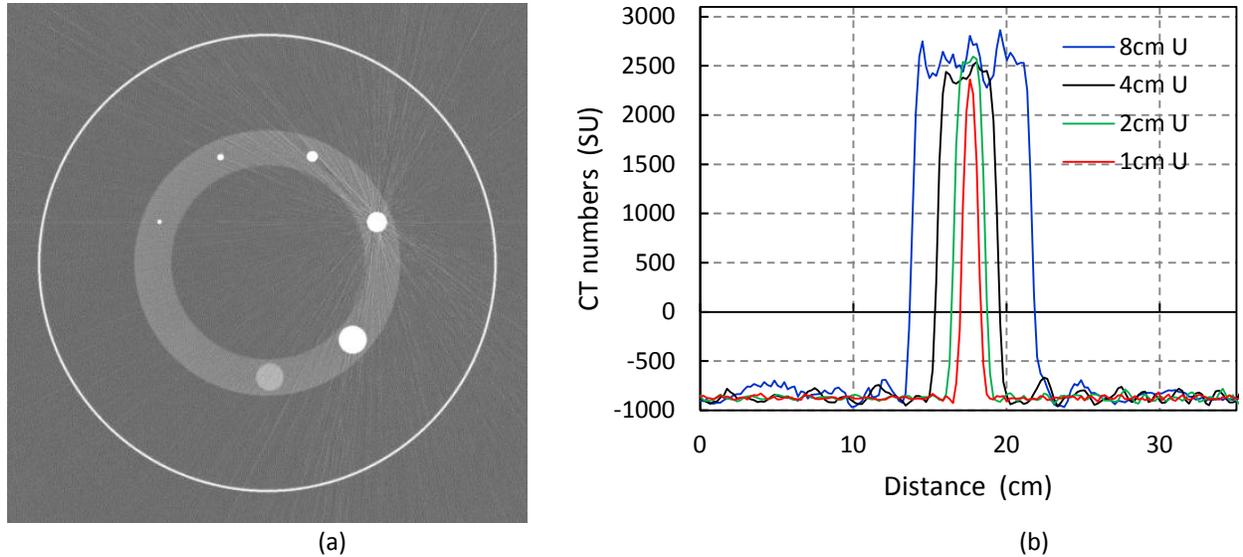

(a) (b)

Figure 9. The MV CT image of the cargo phantom including U spheres with 1cm, 2cm, 4cm, and 8cm diameters (a), and image profiles of the U spheres with different diameters.

The streak artifacts have been further evaluated to clarify how the x-ray dose and attenuation (photon statistics) affect the magnitude of these artifacts. The **figure 10** compares the CT images acquired at 0.2Gy/min and 20Gy/min dose rates, and also at 0.2Gy/min dose rate but with materials having a hollow spherical shapes.

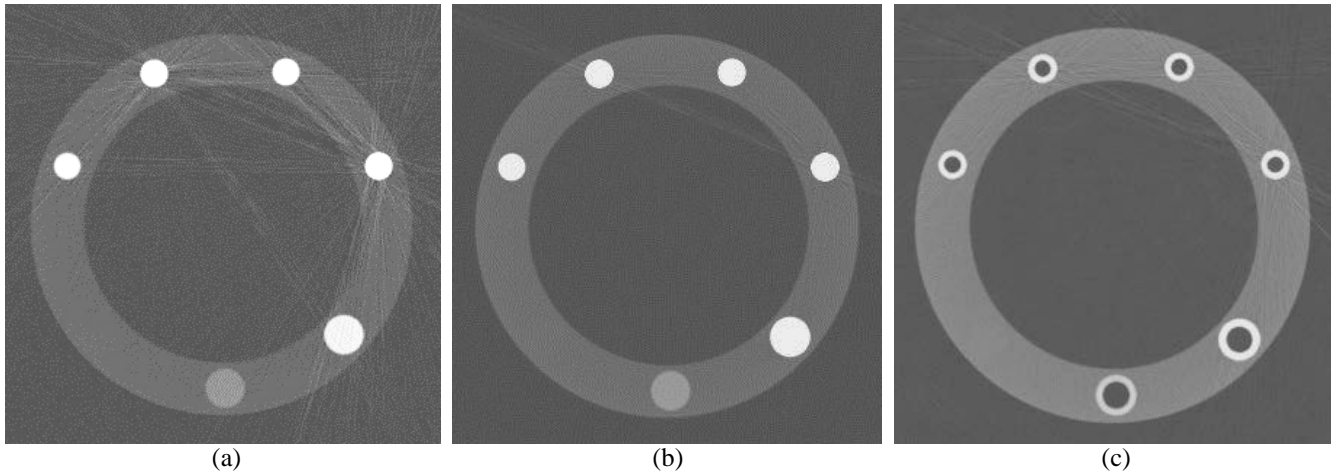

(a) (b) (c)

Figure 10. The CT images of the phantom acquired at 9MV and 0.2Gy/min (a) and 20Gy/min (b) dose rates indicating how streak artifacts due to photon starvation effects depend on dose rate (and photon statistics). The images of the phantom with hollow contrast elements acquired at 9MV and 0.2Gy/min dose rate exhibit lower artifacts due to the lower beam attenuation (c).

It is clear that increasing x-ray dose rate and decreasing x-ray attenuation by using hollow spheres substantially decrease streak artifacts, which is associated with improved photon statistics.

**Figure 11** shows how one of the two materials with different atomic numbers can be cancelled out from the CT image by dual energy subtraction method [19] using the CT images acquired at two energies (in this case at 6MV and 9MV). The **Figure 11** shows the material decomposed images where Polyethylene, steel, and Pb were cancelled out.

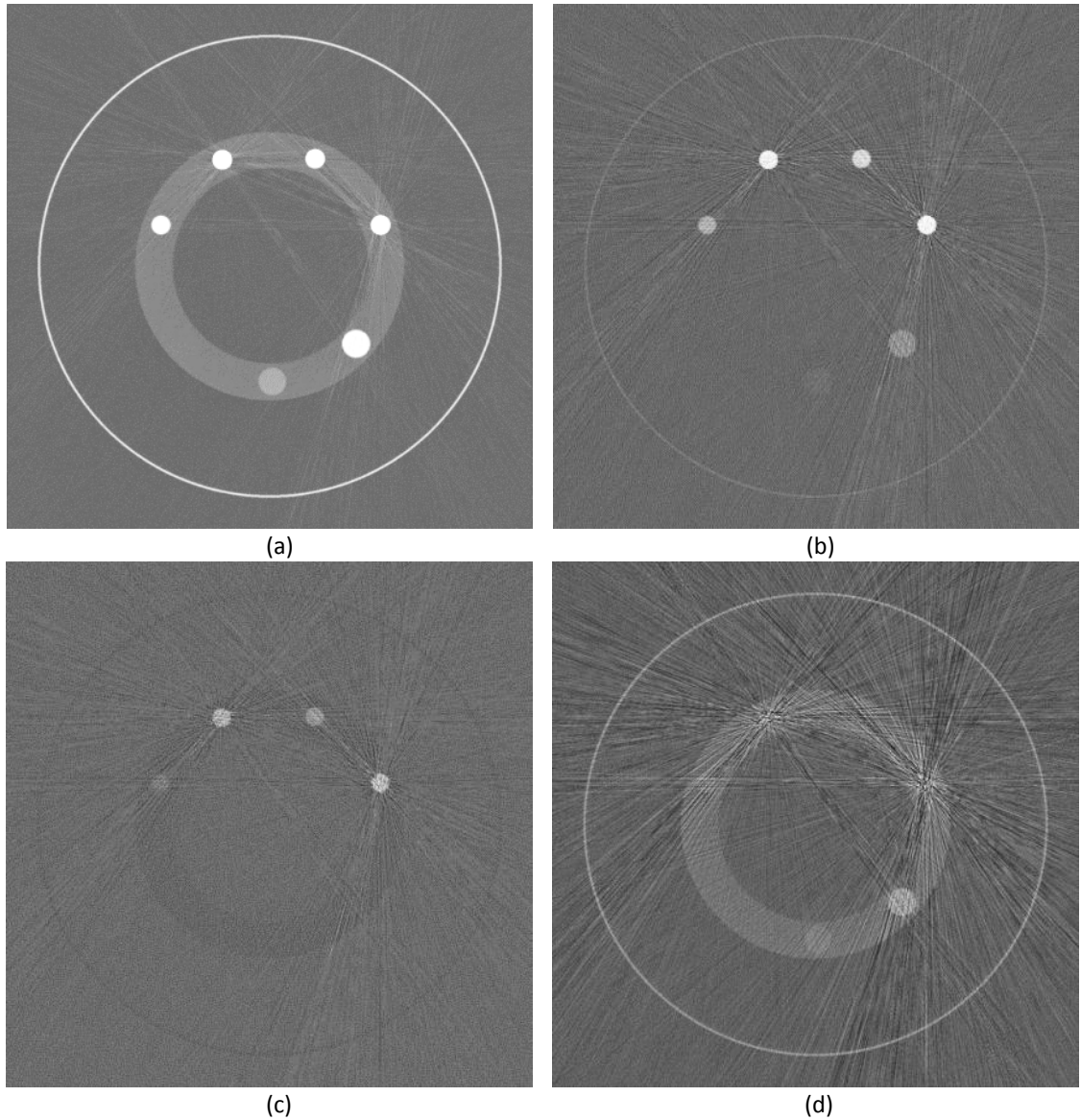

Figure 11. Material decomposed images of the cargo phantom. (a)Original image, (b)Polyethylene cancelled image, (c)Steel cancelled image, and (d)Pb cancelled image.

The signals from other materials remaining after subtraction qualitatively show how well the materials can be separated one from others. For example, all materials except water can be visualized after cancelling out Polyethylene, which means that water and Polyethylene are not separable with this method. Similarly, after cancelling out lead the U and W elements are invisible.

**Figure 12** compares radiography and CT images of the piles of the tubes made from Fe, Pb, W, and U. While the radiography image exhibits cluttered and unrecognizable textures of the tube walls the CT image provides valuable cross-sectional information. On the other hand, the large number of tubes in the pile results in strong x-ray attenuation, elevated noise, and streak artifacts (**figure 12a**). The image quality is substantially improved when the number of tubes in the pile is decreased from 19 to 6 (**figure 12b**).

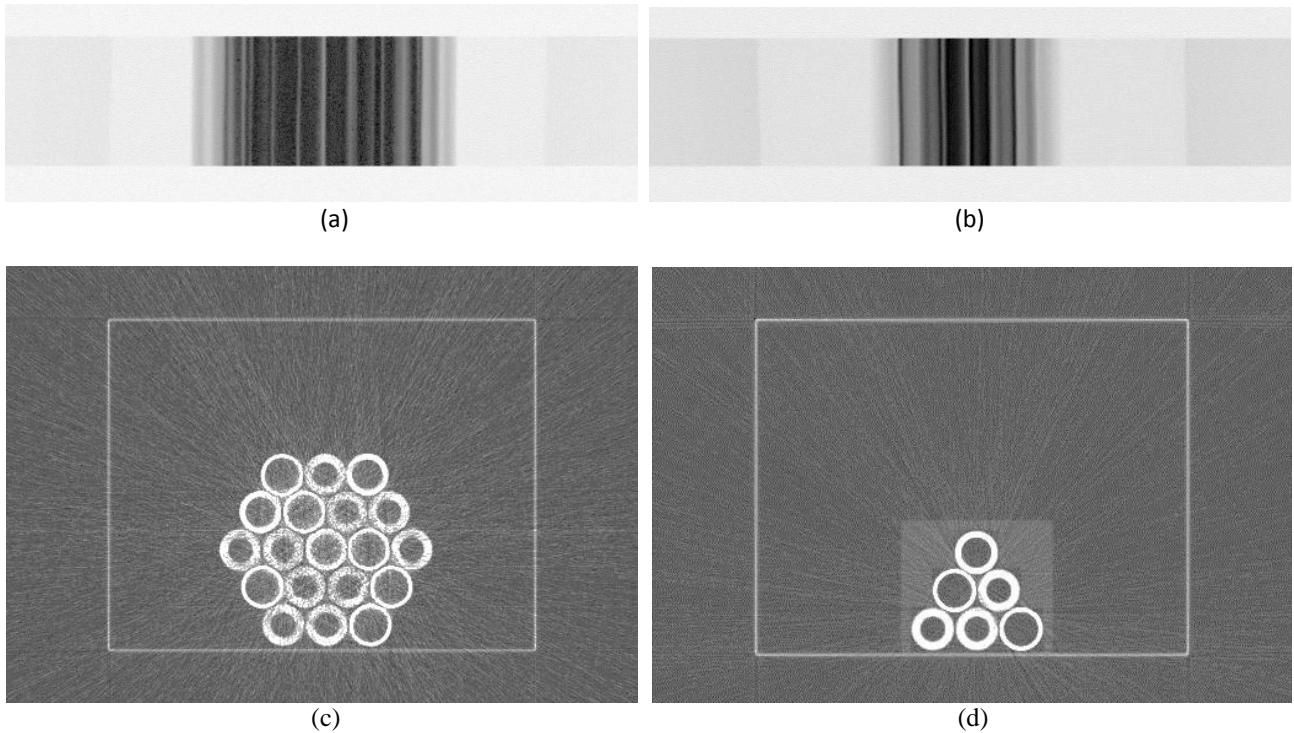

Figure 12. Radiography images of the piles of the tubes made of different materials (a, b). CT images (b, c) of the same tube piles showing cross sectional structure that were not clear in radiography images. The radiography and CT images were acquired with 9MV x-ray beam at 0.8mGy dose.

**4. Discussions and conclusion**

The feasibility study of the large scale MV cargo CT system with 3.25m field of view showed that such a system can be developed and used for imaging large cargos. The system can be developed using compact x-ray sources and detectors that are commercially available and used in existing cargo radiography systems. Although compact x-ray sources provide limited x-ray outputs, their outputs are still sufficient to perform a single- or few-slice CT imaging of cluttered cargo contents for which radiography may be inconclusive. The scans can be performed in 60sec. at the x-ray dose levels that are lower than the permitted dose limit of 5mGy. The materials with close densities and x-ray attenuations such as W and U can be reliably separated and quantified based on their measured CT numbers.

Although acquisition of few CT slices would not cover entire volume of the cargo in axial direction, it still could be very useful for resolving cluttered contents and determining material types based on cross-sectional images and measured CT numbers. To acquire several CT slices at a time, the electron beam can be deflected using a varying magnetic field to hit over the target that is extended in axial direction providing several x-ray focal spots. Each of these focal spots in conjunction with the same detector arc will generate a separate fan beam for acquisition of independent CT slices. Similar target concept has already been used in some MV radiography systems [27].

The MV cargo CT system can perform also single- or multiple-view radiography scans. However, these radiography scans will be slow due to the low x-ray dose output of the compact x-ray sources. Currently, the cargo radiography systems use massive and large x-ray sources with up to 30Gy/min dose outputs that allow for scanning at 60cm/sec. linear speeds of cargo. Therefore, at the current level of source technology the MV cargo CT should be adjunct to the MV radiography system.

Current simulations accounted for only statistical noise due to the photon statistics accumulated in the detector pixels. No electronics noise was included because electronics noise is system dependent and is not disclosed in literature. However, the magnitude of the electronics noise can be estimated indirectly. It is known that for very thick objects the x-ray beam is attenuated so strongly that the magnitude of the remaining x-ray signals becomes comparable or lower than the electronics noise. For example, in MV radiography the largest object thickness that can be imaged (called maximum penetration depth) is approximately 40cm steel equivalent for 9MV beam [4]. Assuming that the highest x-ray output for the stationary 9MV source is approximately 30Gy/min [18], corresponding photon statistics will be approximately $3 \times 10^6$ photon/pixel [19]. Taking into account that 40cm steel attenuates the 9MV x-ray beam by a factor of approximately $1.3 \times 10^5$, the photon statistics after attenuation by 40cm steel is 23photons/pixel. This photon statistics can be used as estimate for detectability limit in the presence of electronics noise. Assuming that MV cargo CT uses the same detector technology as used in MV radiography, the detectability limit of 23photons/pixel can be applied also in MV cargo CT. Because the x-ray output in MV cargo CT is lower than in MV radiography (0.2Gy/min against 30Gy/min), and exposure time is longer (0.083seconds/projection against 0.005sec./line) the pixel statistics in MV cargo CT is $2 \times 10^5$photons/pixel which is by 15 times less than in MV radiography. Therefore the maximum penetration depth in MV cargo CT is also smaller than in radiography and equals to 32cm steel equivalent, which corresponds to approximately by $10^4$ times attenuation. The **figure 13** shows the x-ray attenuation factors for 9MV beam for different materials and material thicknesses. This figure allows for determining maximum thickness of a particular material or combination of several materials that can be imaged without having photon starvation artifacts.

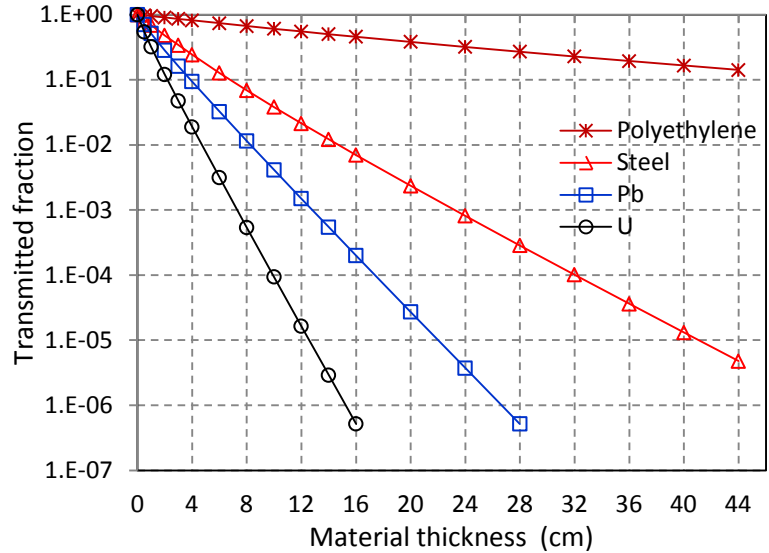

Figure 13. Transmitted fraction of the x-ray photons for 9MV beam plotted against absorber thickness for steel, Pb, and U absorbers.

The streak artifacts observed in CT images presented in the current study are also due to photon starvation effect. However, they are associated with data processing noise inherent to CT reconstruction. Although magnitude of this noise is much smaller than the electronics noise, it created streak artifacts because the photon attenuation in some areas of our phantoms was extremely high resulting in very low photon statistics that are much lower than 23photons/pixel. For example, the x-rays passing the centers of the U and W spheres in **figure 6a** are attenuated by a factor of $5 \times 10^6$, and corresponding photon statistics decreases to 0.04photon/pixel which creates severe strake artifacts. Nevertheless, presence of the streak artifacts did not create a major problem for quantitative assessments of the materials and material separations based on CT numbers, including also separation of W and U with very close x-ray attenuations. Furthermore, in some occasions the above streak artifacts might serve as a useful signature indicating presence of extremely dense high-Z materials such as U and Pu.

Based on the findings of this feasibility study, it is concluded that a large scale MV cargo CT system with 3.25m field of view can be developed based on existing technology that is already used in MV cargo radiography. Such a CT system can be extremely useful as adjunct to existing MV radiography on resolving the cases that cannot be resolved conclusively with radiography alone.